\documentclass[12pt]{iopart}

\usepackage{iopams}
\usepackage{color}
\usepackage{soul}

\usepackage{graphicx} 
\newcommand {\be}{\begin{equation}}
\newcommand {\ee}{\end{equation}}
\newcommand {\ba}{\begin{eqnarray}}
\newcommand {\ea}{\end{eqnarray}}
\newcommand{\Jm}{J_{\mathrm{m}}}
\newcommand{\Jav}{J_{\mathrm{av}}}
\newcommand{\Om}{\omega_{\mathrm{m}}}
\newcommand{\Os}{\omega_{\mathrm{l}}}
\newcommand{\kl}{k_{_L}}
\newcommand{\kr}{k_{_R}}
\newcommand{\kc}{k_{_C}}
\newcommand{\ko}{k_{_0}}
\newcommand{\Vl}{V_{_L}}
\newcommand{\Vr}{V_{_R}}
\newcommand{\nc}{n_{_c}}
\newcommand{\Jl}{J_{_L}}
\newcommand{\Jr}{J_{_R}}
\newcommand{\jif}{J_i^{(1)}}
\newcommand{\jis}{J_i^{(2)}}

\begin{document}


\title[]{Thermal resonance in harmonically driven segmented Frenkel-Kontorova lattices with next-nearest-neighbor interactions}

\author{M.~Romero-Bastida and Ana~Gabriela Mart\'{\i}nez-Rosas}
\address{SEPI ESIME-Culhuac\'an, Instituto Polit\'ecnico Nacional, Av. Santa Ana No. 1000, Col. San Francisco Culhuac\'an, Delegaci\'on Coyoacan, Distrito Federal 04440, Mexico}
\ead{mromerob@ipn.mx}

\date{\today}

\begin{abstract}
Problems of heat transport are ubiquitous to various technologies such as power generation, cooling, electronics, and thermoelectrics. Within this context it is natural that external heat flux control on nanoscale devices became an appealing strategy that has been explored in recent years. In this work we study the thermal resonance phenomenon, i.e., the maximum heat flux obtained by means of an external periodic driving, of a one-dimensional system composed of two dissimilar Frenkel-Kontorva lattices with both nearest-neighbor (NN) and next-nearest-neighbor (NNN) interactions connected by time-modulated NN and NNN couplings in contact with two heat reservoirs operating at different temperature. We study the effect of the NNN interactions on the various heat transport regimes afforded by the structural modifications that can be made on the model. The dependence of the thermal resonance on the system size is studied as well. Our results show that, despite the increased connectivity of both sides afforded by the NNN interactions, the overlap of the phonon bands of both parts of the system still determines the frequency range wherewith thermal resonance emerges. 
\end{abstract}

\noindent{Keywords: transport processes; heat conduction; coupled Frenkel-Kontorova lattices}

\pacs{44.10.+i; 05.60.-k; 05.45.-a; 05.10.Gg}

\maketitle

\section{Introduction\label{sec:Intro}}

Down to the nanoscale thermal management in low dimensional systems is a topic of significant technological importance~\cite{Cahill03,Luo13,Cahill14}. With the emergence of single-molecule electronics~\cite{Cuevas10,Xuefeng14}, where long-chain molecules attached to tiny electrodes are used to transport and switch electrons, the related problem of thermal transport through molecular junctions has received considerable theoretical as well as experimental attention~\cite{Dubi11,Segal16}. It is known that localized Joule heating presents a serious problem for the functionality and reliability of such devices~\cite{Wang07b}. The combination of small molecular heat capacity ---implied by its small size--- and inefficient heat transfer away form it may cause a large temperature increase that will affect the stability and integrity of the molecular junction~\cite{Huang07,Tsutsui08,Tsutsui19}. The rate at which heat is transported away from the conducting junction is, therefore, crucial to the successful realization of nano electronic devices.

In order to address some of the above mentioned issues it has been proposed to employ a temporal modulation to direct heat from one part of the device to another or to a thermal reservoir by means of an applied external work. To test this strategy the use of a simplified model has been of great help in order to reduce the complexity of the problem without sacrificing physical plausibility. As a concrete example, the so-called Frenkel-Kontorova (FK) model~\cite{Braun04} ---which is a one-dimensional (1D) harmonic lattice affected by an onsite potential--- is an attractive option since the coupling of lattices with different parameters indeed affords a simplified model of a molecular junction. In a previous paper we have considered a 1D model consisting of two dissimilar FK lattices connected together by a time-modulated harmonic coupling under the influence of a static thermal bias and determined that the overlap or separation of the phonon bands associated to the dissimilar segments determines the appearance or absence of {\it thermal resonance}, i.e., the maximization of a heat flux directed mainly towards the thermal reservoirs obtained for a specific frequency of the external periodic driving~\cite{Romero20}. Thus this proposal offers a first step to solve the problem of carrying away the heat from the junction between two dissimilar systems. Previously the thermal transport properties of a harmonic lattice system consisting of two semi-infinite leads at equal or different temperature and connected by a time-modulated coupling were studied by means of the nonequilibrium Green's function formalism~\cite{Cuansing10}. Alternate proposals, mainly within the realm of oscillator systems, for external heat flux control involved the time periodic shuttling of one~\cite{Li08} or two~\cite{Ren10} heat reservoirs to obtain a finite heat flux in the absence of a thermal bias. Also, moving barriers have been employed to pump phonons from a cold reservoir to a hotter one~\cite{Arrachea2012}. Heat pumping has been analyzed between semi-infinite harmonic chains with a time-dependent coupling~\cite{Agarwalla11} and ---within the quantum realm--- in anharmonic molecular junctions in contact to phononic baths~\cite{Ren10a}.

In this paper we explore the effect on thermal resonance phenomenon in the aforementioned model of the inclusion of next-nearest-neighbor (NNN) interactions. With this structural modification it becomes feasible to study in a systematic way the effect on thermal resonance ---or any other thermal phenomenon--- of forces with significant magnitude beyond the nearest-neighbor (NN) range. The importance of this possibility is highlighted by the recently reported computation of the relative contributions of NN, NNN, and higher-order interactions to the overall energy flow in molecular junctions composed of alkanedithiol molecules and gold substrates~\cite{Sharony20}. An additional motivation stems from the fact that the thermal transport properties of systems with such a type of interactions are significantly altered by their presence. For example, in a 1D oscillator lattice with randomly distributed NNN interactions ---being thus a model of a glass--- it has been determined that the observed energy localization increases the relaxation time of the system~\cite{RomeroArias08}. Furthermore, the presence of NNN interactions can change the size dependence of the thermal conductivity when the heat-carrying phonons are scattered by a quartic anharmonicity~\cite{Xiong12,Xiong14}. And finally, the asymmetric heat flow, i.e., thermal rectification, of some 1D oscillator systems has been shown to improve by the addition of those same interactions~\cite{Romero21}. Therefore, it is reasonable that there might be an appreciable influence of the NNN interactions on the thermal resonance phenomenon of the the herein considered system.

The remainder of the paper is organized as follows: in Sec.~\ref{sec:Model} the model system and methodology are presented. Our results on the dependence of the thermal resonance on the structural parameters of the model in the presence of NNN interactions are reported in Sec.~\ref{sec:Res}. The discussion of the results, as well as our conclusions, are presented in Sec.~\ref{sec:Disc}.

\section{The Model\label{sec:Model}}

\begin{figure}
\centerline{\includegraphics*[width=95mm]{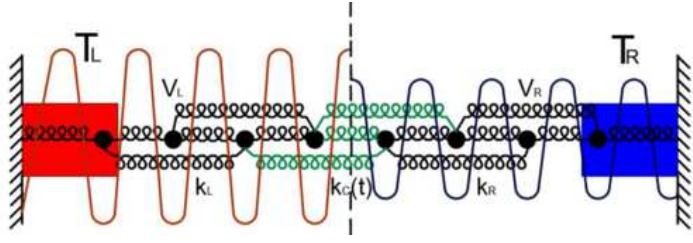}}
\caption{Schematic representation of two dissimilar FK lattices with both NN and NNN interactions connected by a time-modulated harmonic interaction. The whole system is attached at both ends to a thermal reservoir, each operating at different temperatures.}
\label{fig:1}
\end{figure}

The herein considered system consists of a size $N$ oscillator lattice that includes both NN and NNN interactions consisting of two segments ($L,R$) of size $\nc=N/2$ coupled together by harmonic springs with a time-modulated strength $\kc(t)$; a sketch of the model is presented in Fig.~\ref{fig:1}. The equations of motion (EOM) of any given oscillator within each segment can be written, in dimensionless variables, as $\dot q_i =p_i/m_i$ and 
\ba \dot p_i& = &F^{(1)}_i + \gamma F^{(2)}_i - {V_{_{L,R}}\over2\pi}\sin(2\pi q_i/a) \cr
   & + & (\xi_{_1} - \alpha_{_1} p_i)\,\delta_{i1} + (\xi_{_N} - \alpha_{_N} p_i)\,\delta_{i{N}},
\ea
being $F^{(1)}_i=F(q_i-q_{i-1})-F(q_{i+1}-q_i)$ and $F^{(2)}_i=F(q_i-q_{i-2})-F(q_{i+2}-q_i)$, where $F(x)=-\partial_x V(x)$ with $V(x)=k_{_{L,R}}x^2/2$. $\{m_i,q_i,p_i\}_{i=1}^{N}$ are the dimensionless mass, displacement, and momentum of the $i$th oscillator. The tunable parameter $\gamma$ specifies the relative strength of the NNN interaction $F^{(2)}_i$ compared to the NN one $F^{(1)}_i$. $k_{_{L,R}}$ and $V_{_{L,R}}$ are the harmonic spring constant and the amplitude of the FK onsite potential in each segment, respectively. Just as done in previous works~\cite{Li04a,Romero20} we set $\Vr=\lambda\Vl$ and $\kr=\lambda\kl$ in order to reduce the number of adjustable parameters. The above EOM corresponds to the commensurate case ---which is the only one considered in this study--- where the onsite potential assumes the same spatial periodicity as the lattice constant $a=1$. Henceforth we will consider a homogeneous system, i.e., $m_i=1\,\,\forall\,i$ and fixed boundary conditions ($q_{_0}=q_{_{{N}+1}}=0$). The stochastic force $\xi_{_{1,N}}$ is a Gaussian white noise with zero mean and correlation $\langle\xi_{_{1,N}}(t)\xi_{_{1,N}}(t^{\prime})\rangle=2\alpha_{_{1,N}}T_{_{1,N}}m_i(\delta_{1i}+\delta_{{N}i})\delta(t-t^{\prime})$, being $\alpha_{_{1,N}}$ (taken as $0.5$ in all computations hereafter reported) the coupling strength between the system and the left and right thermal reservoirs operating at temperatures $T_{_L}=0.15$ and $T_{_R}=0.05$, respectively; the system thus operates at a constant average temperature value of $T_{_0}\equiv(T_{_L}+T_{_R})/2=0.1$. The external periodic drive with frequency $\omega$ will affect the NN forward interaction between $q_{_{\nc+1}}$ and $q_{_{\nc}}$, as well as the NNN one of $q_{_{\nc}}$ with $q_{_{\nc+2}}$ and the backward one of $q_{_{\nc+1}}$ with $q_{_{\nc-1}}$; these effects are taken into account by means of a harmonic coupling with a time-modulated amplitude given by $\kc(t)=\ko(1+\sin\omega t)$, as sketched in Fig.~\ref{fig:1}. This choice corresponds to a rough approximation of the effect of the external drive acting on the contact region, i.e. the space between oscillators $q_{_{\nc}}$ and $q_{_{\nc+1}}$, wherein both NN and NNN interactions that afford a connection between both lattices exist. Experimentally this could be achieved in molecular junctions by varying the distance among them by means of a highly localized external field acting upon the contact region. The complete set of EOM were integrated with a stochastic velocity-Verlet integrator with a timestep of $10^{-2}$ for a production time interval of $2\times10^7$ time units after a transient time of $10^{8}$ time units.

Once the non-equilibrium stationary state is attained, the local heat flux is computed as
\ba
J_i& =\langle\dot q_{i} F(q_{i+1}-q_i)\rangle + 2\gamma\langle\dot q_{i+2} F(q_{i+2}-q_i)\rangle \nonumber \\
   & =\jif + 2\gamma\jis \label{lhf}
\ea
and the local temperature as $T_i=\langle p_i^2/m_i\rangle$; in both instances $\langle\cdots\rangle$ indicates time average. In the stationary state the heat flux in each segment becomes independent of the site, and, in order to improve the statistical precision of our results, the mean heat flux $J_{_{L,R}}$ on each side of the lattice is calculated as the algebraic average of ${J}_i$ over the number of unthermostatted oscillators in each segment. Henceforth we will hold the convention that $J_{_{L,R}}$ are positive when heat flows into the reservoirs and negative otherwise. Thus the average heat flux $\Jav\equiv(\Jl+\Jr)/2$ is proportional to the power supplied by the external driving and dissipated into the reservoirs.

\section{Results\label{sec:Res}}

\begin{figure}\centering
\centerline{\includegraphics*[width=95mm]{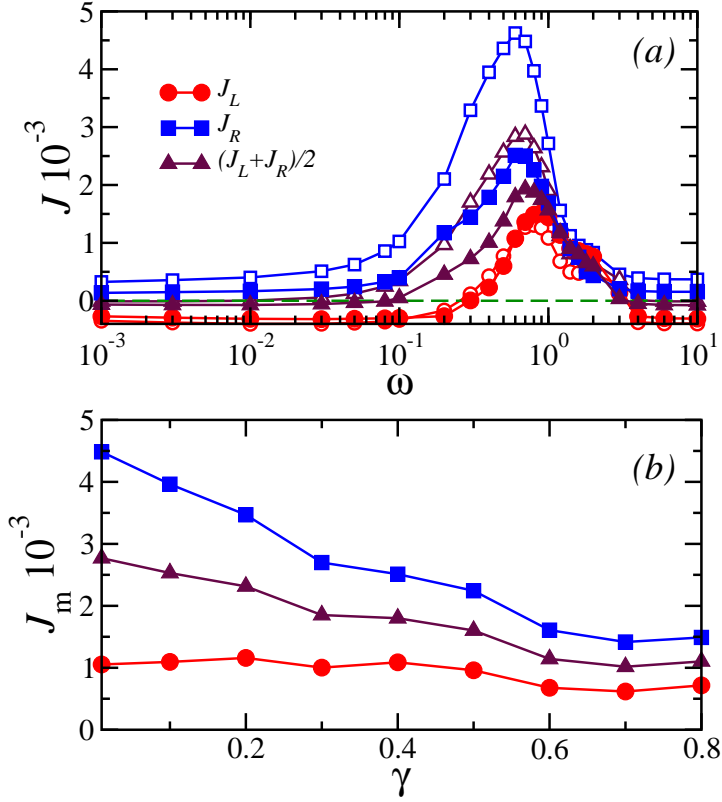}}
\caption{(a) Heat flux vs driving frequency $\omega$ for $\gamma=0.4$. The energy currents through the left and right segments are $\Jl$ and $\Jr$, respectively, with $J$ being the average current. Void symbols correspond to the results for $\gamma=0$ of Ref.~\cite{Romero20}. (b) Maximum heat flux on the left (circles) and right (squares) segments vs relative strength of the NNN interactions $\gamma$, together with their average (triangles); in all instances thermal resonance occurs at $\Om=0.7$. In both panels $\Vl=5$, $\kl=1$, $\lambda=0.2$, $\ko=0.05$, $N=32$, and $\nc=N/2$. Continuous lines are a guide to the eye.}
\label{fig:2}
\end{figure}

Just as in the NN case already studied~\cite{Romero20}, within a defined frequency range the resonant interaction of the external drive with the system's intrinsic frequencies leads to the thermal resonance phenomenon wherewith the power released by the external drive in the contact is dissipated into the reservoirs. To study this result in more detail the behavior of the heat fluxes $\Jl$ and $\Jr$ along each segment of the system as a function of the driving frequency $\omega$ is presented in Fig.~\ref{fig:2}(a) for a relative strength of the NNN interactions of $\gamma=0.4$ with $\Vl=5$, $\Vr=1$ $\kl=1$, $\kr=0.2$, $\ko=0.05$, and $N=32$. The results obtained in Ref.~\cite{Romero20}, corresponding to $\gamma=0$ with the aforementioned parameter values, are also presented for comparison. As expected, both in the adiabatic driving limit $\omega\rightarrow0$ and in the opposite one $\omega\rightarrow\infty$ the heat transport is dominated by the imposed temperature gradient, which results in $\Jr=-\Jl>0$, and thus the average heat flux $\Jav=0$ with no power dissipated in the contact region whatsoever; in both limits the system reduces to two segments interacting via a constant harmonic coupling. Within the intermediate frequency range of $0.26\lesssim\omega\lesssim2$ thermal resonance sets in, since now $J_{_{L,R}}>0$, and the average heat flux attains a maximum value $\Jm$ at $\Om=0.7$, as already mentioned. It can also be observed that the heat flux across the right segment for $\gamma=0.4$ is lower than the corresponding value for $\gamma=0$ in all the considered frequency range, whereas the flux across the left one is very similar for both $\gamma$ instances considered. The dependence of the maximum heat flux $\Jm$ attained within each segment with respect to the relative strength of the NNN interactions $\gamma$ is presented in Fig.~\ref{fig:2}(b); in all $\gamma$ instances the maximum heat flux is attained at the same resonant frequency of $\Om=0.7$. It can be readily appreciated that $\Jm$ steadily declines, specially in the right segment, as $\gamma$ increases, an effect due to the increased coupling between both segments afforded by the NNN interactions in the contact region. It is also noted that the $\gamma=0.4$ value affords an adequate description of the system's behavior for not too low $\gamma$ values that would render the same results as $\gamma=0$ and not too high values wherewith the thermal resonance is greatly diminished. Therefore, unless otherwise stated, we will hereafter employ $\gamma=0.4$ as a representative value to gauge the influence of the NNN interactions on thermal resonance.

\begin{figure}
\centerline{\includegraphics*[width=95mm]{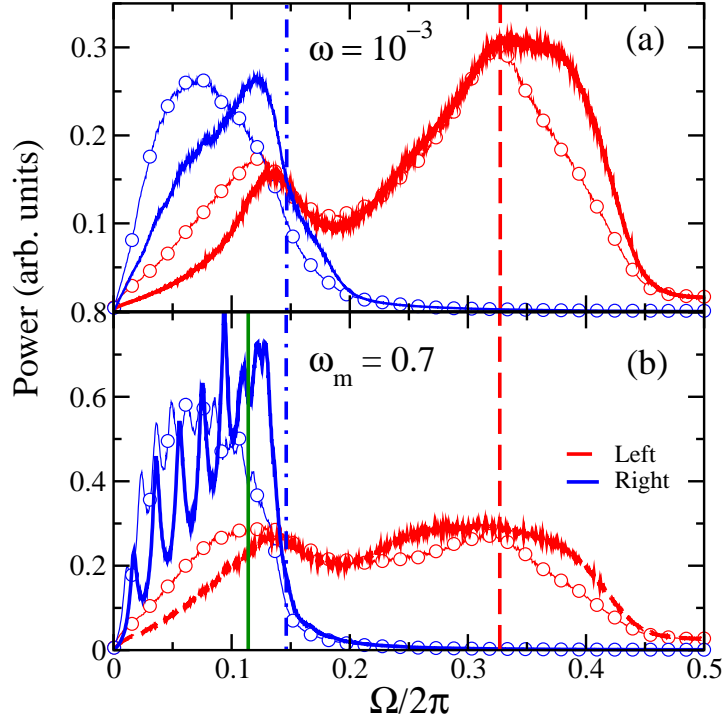}}
\caption{(a) Power spectra of the two oscillators in the left (red) and right (blue) sides of the contact for a driving frequency of $\omega=10^{-3}$ and $\gamma=0.4$. (b) Same as (a) but for $\Om=0.7$. Vertical dashed and dot-dashed lines correspond to the cut-off frequencies of the left and right phonon bands respectively. Vertical solid line denotes the $\Om/2\pi$ value. Same $\Vl$, $\kl$, $\lambda$, and $\ko$ values as in Fig.~\ref{fig:2}. Void symbols correspond to the results for $\gamma=0$ of Ref.~\cite{Romero20}.}
\label{fig:3}
\end{figure}

To study in more detail the phenomenology due to the contribution of the NNN interactions in Fig.~\ref{fig:3} we plot the power spectra $|\tau^{-1}\!\!\int_{_0}^{\tau}\!\! dt\dot q_i(t)\exp(-\mathrm{i}\Omega t)|^2$ of the interface oscillators $i=16,17$ at the left and right sides of the contact for both the adiabatic driving limit, $\omega=10^{-3}$, and in the thermal resonance regime, $\Om=0.7$, which corresponds to the maximum average heat flux $\Jm$ of Fig.~\ref{fig:2}(a). In the former limit $T_{16}=0.1402$ and thus $T_{_L}-T_{16}\sim\mathcal{O}(10^{-3})$; furthermore, $T_{17}=0.0577$ and $T_{17}-T_{_R}\sim\mathcal{O}(10^{-3})$. Thus, since the difference between the temperature values in the boundaries and those in the interface is not significant, it is possible to consider $T_{_L}$ as a representative temperature for the left side ---or an upper bound to the temperatures in that same side--- and $T_{_R}$ as the lower bound of the local temperatures in the right one. Now, both sides of the system are in a temperature regime above a threshold value $T_{_{\mathrm{cr}}}\approx V/(2\pi)^2$ in which they behave as harmonic lattices, since $T_{_{\mathrm{cr}}}^{(L)}=0.13<T_{_L}$ for $V_{_L}=5$ and $T_{_{\mathrm{cr}}}^{(R)}=0.025<T_{_R}$ for $V_{_R}=1$. This fact is further corroborated by noticing that, if the system were in the anharmonic regime, the lower bound of the phonon bands would be raised by $\sqrt{V_{_{L,R}}}$~\cite{Li04a}, which would prevent any phonons existing at all in the frequency range $\Omega/2\pi<0.0566$ in the left side and in $\Omega/2\pi<0.1592$ in the right one, and this is clearly not the case for the phonon bands displayed in Fig.~\ref{fig:3}. In the case where NNN interactions are present the dispersion relation reads as $\omega_{\alpha}=2[(k_{_{L,R}}/m_{_{L,R}})(\sin^2q_{\alpha}/2+\gamma\sin^2q_{\alpha})]^{\frac{1}{2}}$, where $q_{\alpha}$ is the wave number and $\omega_{\alpha}$ the corresponding frequency; the maximum value $\omega_{\mathrm{max}}$ for a certain wave number indicates the phonon frequency limit for the considered side of the system, which define the phonon band $0<\Omega<2\omega_{\mathrm{max}}$. Therefore, for $\gamma=0.4$ we have $\omega_{\mathrm{max}}/2\pi\sim0.3274$ for $\kl=1$ and $\omega_{\mathrm{max}}/2\pi\sim0.1466$ for $\kr=0.2$ respectively, which define the phonon band $0<\Omega/2\pi\lesssim0.3274$ for the left oscillator and $0<\Omega/2\pi\lesssim0.1466$ for the right one. Within this framework we can readily see that for $\omega=10^{-3}$ there are frequency values of both spectra in the overlapping region $\Omega/2\pi<0.1466$ of these phonon bands, a feature that allows a heat transport regime dominated by the imposed thermal bias~\cite{Li04a}. For the $\Om=0.7$ case presented in Fig.~\ref{fig:3}(b) we have $T_{16}=0.184$ and $T_{17}=0.124$, thus $T_{16}\gg T_{_{\mathrm{cr}}}^{(L)}$ and $T_{17}\gg T_{_{\mathrm{cr}}}^{(R)}$; therefore in this regime is more natural to consider the temperatures of the interface oscillators as those representative of their respective sides since these oscillators are the ones directly involved in the observed thermal resonance phenomenon. Furthermore, in the thermal resonance regime both sides of the system are well above the corresponding temperature threshold values, and so the same phonon bands for the adiabatic regime can be employed here as well. Both spectra have more than twice the spectral power compared to those in the adiabatic limit, being this increase most evident in the spectrum corresponding to the right side of the system, which in turn accounts for $\Jr>\Jl$. Thus, this increase of the spectral power in the overlapping region, together with the reduction in the high-frequency contribution of the left spectrum compared to the corresponding one in the adiabatic regime presented in the previous panel, is responsible of the thermal resonance phenomenon observed in Fig.~\ref{fig:2}(a). However, we also notice that the contribution of high-frequency phonons in the left side remains larger ---although only marginally--- than that of the low-frequency ones in the driving regime. This fact certainly helps to explain the overall reduction of the $\Jm$ value when NNN interactions are present compared to the case where only NN contributions are relevant, as can be seen in Fig.~\ref{fig:2}(a).

\begin{figure}\centering
\centerline{\includegraphics*[width=95mm]{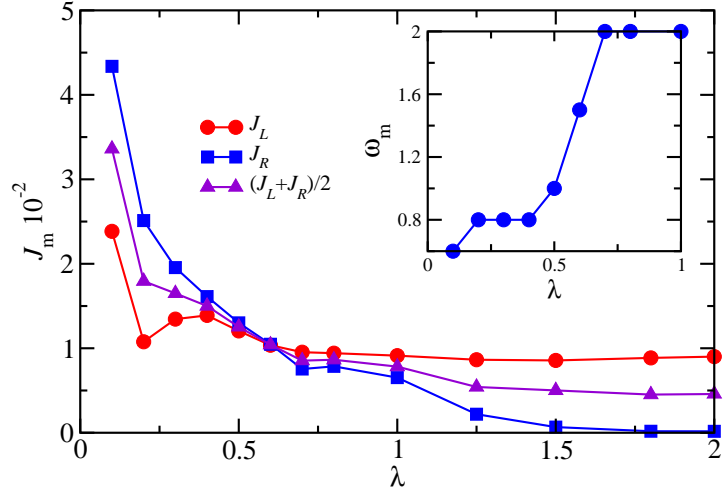}}
\caption{Heat flux $\Jm$ vs scaling parameter $\lambda$ with $\Vl=5$, $\kl=1$, and $\ko=0.05$. Inset is $\Om$ vs $\lambda$. Continuous lines are a guide to the eye.}
\label{fig:4}
\end{figure}

Next we will explore the properties of the system as a function of the scaling parameter $\lambda$, which controls the effects of changes in both the elastic constant and the strength of the onsite potential. In Fig.~\ref{fig:4} we plot the dependence of $\Jm$ versus $\lambda$, wherein it can be seen that, for $\lambda<0.6$, we have $\Jr>\Jl$ and thus more heat flows into the cold reservoir. For larger $\lambda$ values the heat flow into the hot reservoir becomes almost $\lambda$-independent, whereas there is a steady decrease of heat flow into the cold reservoir connected to the right side of the system; within this $\lambda$ value range both the increasing height of the potential and the stiffness of the harmonic constant further contribute to confine the oscillators, which have low kinetic energy due to the low temperature in the right side, in the potential valley. Therefore the heat flux in the right side is greatly diminished and thus results in $\Jl>\Jr$. The more important difference with the NN case previously studied is that the $\lambda$ value at which the aforementioned change of relative magnitude in the heat fluxes occurs is lower than that corresponding to the already known NN case. Thus, our results suggest that a strong thermal resonance effect is restricted to highly asymmetric systems when NNN interactions are present. We can explain this fact by noticing that, as $\lambda\rightarrow1$, the system's symmetry increases, which entails a reduction in the magnitude of the heat fluxes. This reduction in magnitude of the thermal resonance as the system increases its structural symmetry is akin to the reduction of the thermal rectification, i.e., asymmetric heat flow, of this and similar systems under the aforementioned condition~\cite{Li04a}. Now, the increased connectivity afforded by the NNN interactions further contributes to the reduction of the heat fluxes $\Jl$ and $\Jr$, together with the eventual saturation of the latter in a lower value than the former; then the result is that $\Jl>\Jr$ occurs at a lower $\lambda$ value. In the inset we plot $\Om$ versus $\lambda$ and it can be readily observed that, as $\lambda$ increases, the resonant frequency has a sharp increase within a $\lambda$ value range in which the aforementioned change form $\Jr>\Jl$ to $\Jl>\Jr$ was noticed. 

\begin{figure}\centering
\centerline{\includegraphics*[width=95mm]{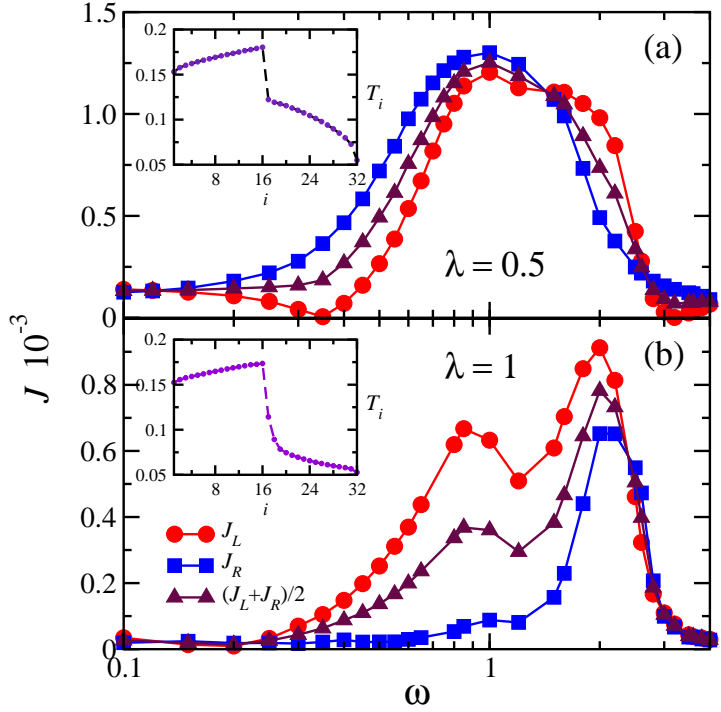}}
\caption{(a) Heat flux vs driving frequency $\omega$ for $\lambda=0.5$. (b) Same as (a) but for $\lambda=1$; the two peaks of the heat flux correspond to frequencies $\Os=0.8$ and $\Om=2$. In the inset of each panel the corresponding temperature profile for the thermal resonant frequency $\Om$ of each case is presented. In both panels $\Vl=5$, $\kl=1$, and $\ko=0.05$. Continuous lines are a guide to the eye.}
\label{fig:5}
\end{figure}

In Fig.~\ref{fig:5} we plot the heat flux as a function of the driving frequency for two $\lambda$ values. In the upper panel it can be seen that the behavior is not very different to that already reported in Fig.~\ref{fig:2}(a), with some differences worth remarking. The resonant frequency $\Om$ is shifted towards a higher frequency value and the curves are more similar to one another than in the $\lambda=0.2$ case depicted in Fig.~\ref{fig:2}(a), an effect caused by both the increase in the symmetry of the system and the presence of NNN interactions that in turn increase the phonon transmission between both connected lattices. For the temperature profile presented in the inset it can be readily appreciated that the jump in the part of the temperature profile in contact with the cold reservoir increases the slope of the curve in the right side compared to the one in the left one, which results in $\Jr>\Jl$ as observed in Fig.~\ref{fig:4}. For the $\lambda=1$ case displayed in the lower panel it is seen that the main peak corresponding to $\Om$ is shifted to an even higher frequency value compared to the $\lambda=0.5$ case and a secondary peak at a frequency $\Os\sim0.8$ is also present. Furthermore, $\Om$ is clearly associated with the heat flux $\Jl$ of the left side of the system, which is certainly compatible with the fact that $\Jl>\Jr$ when $\lambda>0.6$. The slope of the right side of the temperature profile reported in the corresponding inset  drops significantly compared to the $\lambda=0.5$ case, and thus now $\Jl>\Jr$. Since for $\lambda=1$ value the system is completely symmetrical on both sides ---although still weakly connected since $\ko=0.05$---, we can infer that the smoothing of the temperature profile at the interface is a consequence of the presence of the NNN interactions, which further increase the similarity in behavior between both sides of the system.


\begin{figure}
\centerline{\includegraphics*[width=95mm]{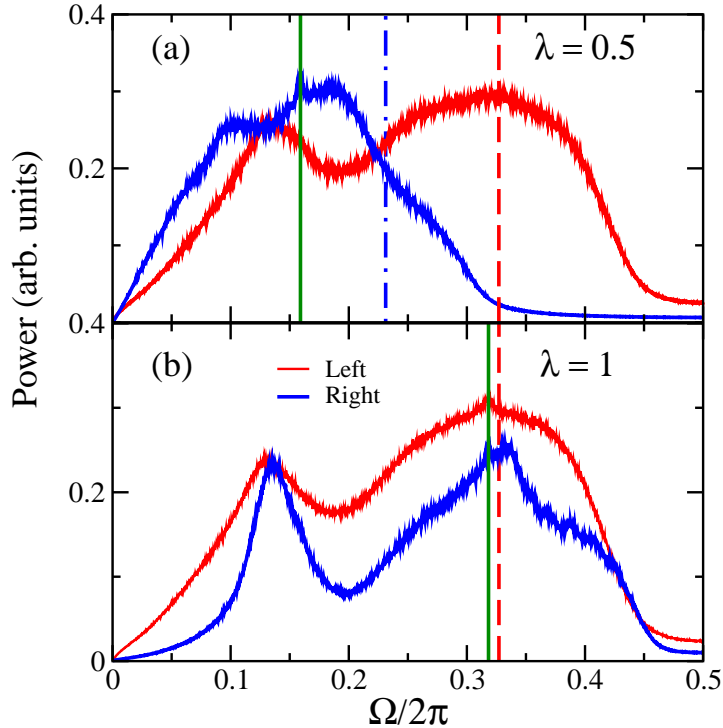}}
\caption{(a) Power spectra of the two oscillators in the left (red) and right (blue) sides of the contact for a $\lambda=0.5$ value. (b) Same as (a), but for $\lambda=1$. Vertical dashed and dot-dashed lines correspond to the cut-off frequencies of the left and right phonon bands in (a) and vertical dashed line in (b) to the cut-off frequency of the left phonon band. See text for further details. Vertical solid line denotes the corresponding $\Om/2\pi$ values in each case. In both panels $\Vl=5$, $\kl=1$, and $\ko=0.05$.}
\label{fig:6}
\end{figure}

The power spectra of the oscillators to the left and right of the contact for the $\lambda=0.5$ value of Fig.~\ref{fig:4} are displayed in Fig.~\ref{fig:6}(a). Compared to the results of Fig.~\ref{fig:3}(b) it is seen that the increase in symmetry in the system has reduced the height of the spectrum corresponding to the right side and has widened it to include higher frequencies. For this $\lambda$ value we have $T_{16}=0.1798\gg T_{_{\mathrm{cr}}}^{(L)}=0.13$ and $T_{17}=0.1221\gg T_{_{\mathrm{cr}}}^{(R)}=0.0633$, which indicates that both sides can again be considered as harmonic lattices. Therefore in this instance the phonon bands are given by $0<\Omega/2\pi\lesssim0.33$ for the left side and $0<\Omega/2\pi\lesssim0.23$ for the right one; the value for which thermal resonance appears, $\Om/2\pi\sim0.159$, coincides with a spike-like value in the spectra for the right side, being the dominant one in the heat conduction as was evident in Figs.~\ref{fig:4} and~\ref{fig:5}(a). For the $\lambda=1$ instance, which is the one associated with maximum symmetry in the system, the qualitative similitude of both spectra is highly increased, as can be readily appreciated in Fig.~\ref{fig:6}(b). The left peak of the $J$ vs $\omega$ plot in Fig.~\ref{fig:5}(b) can be associated with the leftmost ones of the displayed spectra since they are located at $\Omega/2\pi\sim0.13$, a value almost identical to $\Os/2\pi$. For the considered $\lambda$ value now we have $T_{16}=0.1734\gg T_{_{\mathrm{cr}}}^{(L,R)}=0.13>T_{17}=0.1141$, which indicates that now $\Vr$ becomes relevant, with a right phonon band now given by $\sqrt{\Vr}<\Omega<\sqrt{\Vr+4\omega_{\mathrm{max}}^2}$~\cite{Li04a} which, for the considered conditions, is $0.3559<\Omega/2\pi\lesssim0.7448$. The latter has no possible overlap with the left phonon band $0<\Omega/2\pi\lesssim0.33$; thus the latter is the only open channel available for heat carrying phonons. Therefore the resonant frequency, which is $\Om/2\pi\sim0.32$, lies precisely within this phonon band. Furthermore, $\Om$ coincides with the rightmost peaks of both spectra; since the spectra of the left side has more power, the corresponding heat flux is dominant, as was already seen in Fig.~\ref{fig:4}. Now, as previously noticed thermal resonance diminishes as the system's symmetry increases, in a similar way as in thermal rectification for this very same system. In the latter effect structural asymmetry is a fundamental condition for its appearance. But thermal resonance seems to be a more resilient effect, as it can exist under the lack thereof, although greately diminished. However, under maximum symmetry conditions there is a lack of low-frequency overlap of two clearly distinct spectra such as that observed for the highly asymmetric conditions of Fig.~\ref{fig:3}. This situation now allows phonons of all frequencies to be involved in the process. Thus, the main and secondary peaks in the $J$ vs $\omega$ diagram of Fig.~\ref{fig:5}, which appear on the same frequency $\Om$ and $\Os$ values as those in the corresponding power spectrum displayed on Fig.~\ref{fig:6}, indicate the involvement of both high and low energy phonons in the process. The dominance of high-frequency phonons over the low-frequency, heat-carrying ones seems to be responsible of the decrease in magnitude of thermal resonance.

\begin{figure}\centering
\centerline{\includegraphics*[width=95mm]{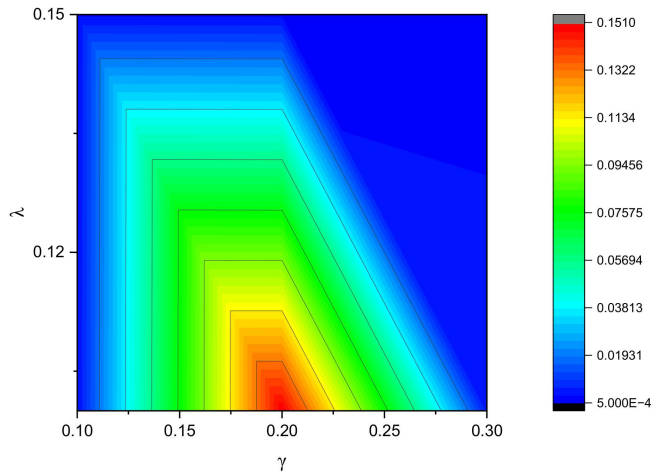}}
\caption{Contour plot $J(\gamma,\lambda)$ for $\Vl=5$, $\kl=1$, and $\ko=0.05$.}
\label{fig:7}
\end{figure}

Now, to complete this part of our study the contour plot of $\Jm(\gamma,\lambda)$ is depicted in Fig.~\ref{fig:7}. The presented data are consistent with our previous results: the presence of NNN interactions constrains the maximum $\Jm$ values to highly asymmetric systems, i.e., those with $\lambda\ll1$. Furthermore, even for $\lambda=0.1$ the maximum average heat flux has low values for $\gamma>0.3$ ---for $\gamma\ll1$ the effect of the NNN interactions becomes negligible, which accounts for the low $\Jm$ value in the aforementioned $\gamma$ range---. Thus, our results indicate the regions of $\lambda$ and $\gamma$ values wherewith the maximum heat flux can be obtained.

\begin{figure}\centering
\centerline{\includegraphics*[width=95mm]{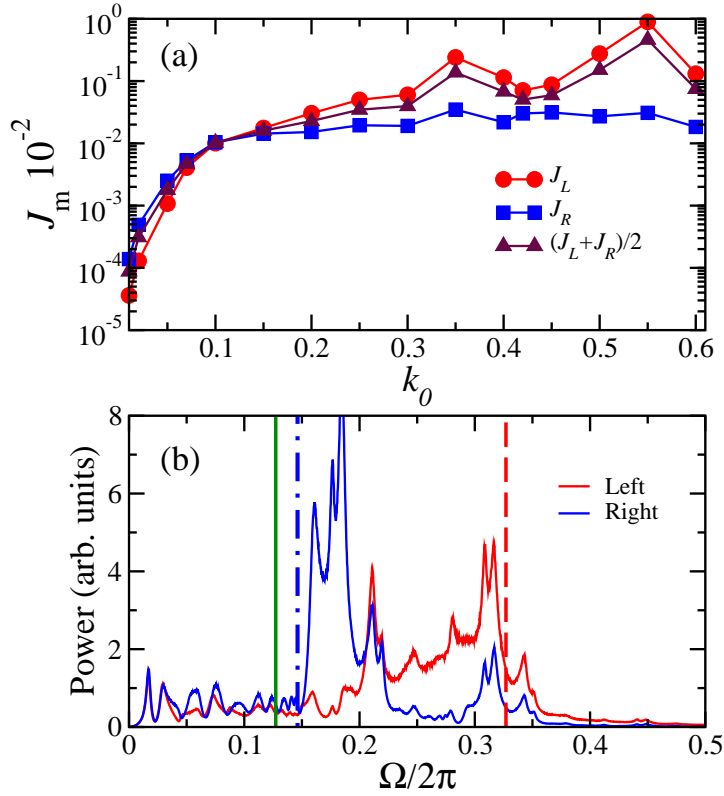}}
\caption{(a) Heat flux $\Jm$ vs $\ko$ with $\Vl=5$, $\kl=1$, and $\lambda=0.2$. Continuous lines are a guide to the eye. (b) Power spectra of the two oscillators in the left (red) and right (blue) sides of the contact for a $\ko=0.6$ value of the interface elastic constant. Vertical dashed and dot-dashed lines correspond to the cut-off frequencies of the left and right phonon bands, respectively. Vertical solid line denotes the $\Om/2\pi$ value.}
\label{fig:8}
\end{figure}

In Fig.~\ref{fig:8}(a) we present the variation of $\Jm$ as a function of the interface elastic constant $\ko$. It can be clearly seen that the regime wherein we have been working so far ---in which there is a higher heat flux into the cold reservoir--- is limited to $\ko<0.1$ values, wherein both heat fluxes increase drastically as $\ko$ approaches the aforementioned limit value. For higher $\ko$ values the maximum heat flux into the cold reservoir becomes at a large extent independent of the value of interface elastic constant whereas the heat flux into the hot reservoir experiences a slow, non-monotonic increment as $\ko$ further increases, and thus $\Jl>\Jr$. The change in relative magnitude of $\Jl$ and $\Jr$ as $\ko$ increases can be explained by inspecting the power spectra corresponding to $\ko=0.6$, a rather high value of the interface elastic constant, that are plotted in Fig.~\ref{fig:8}(b); the left and right phonon bands are $0<\Omega/2\pi\lesssim0.3274$ and $0<\Omega/2\pi\lesssim0.1466$, just as in Fig.~\ref{fig:3}(b). It can be immediately observed that the spectra are more closely entangled in the low frequency region; thus the ensuing resonant frequency $\Om=0.12$ lies within the overlapping region of both spectra, just as in the $\ko=0.05$ case depicted in Fig.~\ref{fig:3}(b). The presence of NNN interactions results in a even closer entanglement of both segments, which is reflected as an increased power available in the spectra ---the same effect as that already observed in the case depicted in Fig.~\ref{fig:3}(b), but of much higher magnitude than in the foregoing one---. However, more power is available at higher frequencies in both spectra of Fig.~\ref{fig:8}(b). Within this frequency range the spectrum associated to the left side has more high frequencies activated, which on average carry more power and thus $\Jl>\Jr$, in agreement with the results of Fig.~\ref{fig:8}(a).

\begin{figure}\centering
\centerline{\includegraphics*[width=95mm]{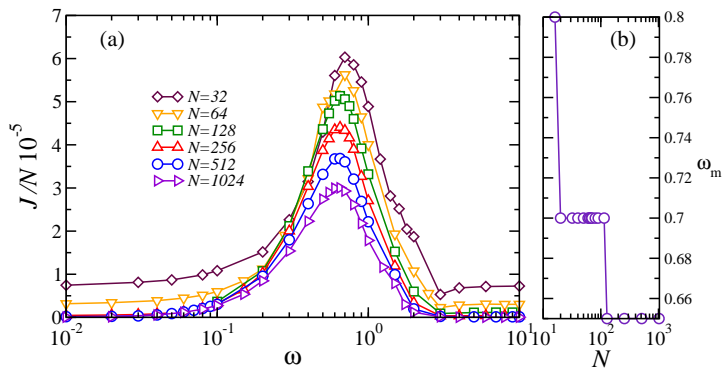}}
\caption{(a) Heat flux vs driving frequency for various system sizes $N$. (b) Resonant frequency vs system size. $\gamma=0.4$, $\Vl=5$, $\kl=1$, $\lambda=0.2$, and $\ko=0.05$ in both panels.}
\label{fig:9}
\end{figure}

By varying $N$ the maximum value of the average heat flux $\Jm$ increases with system size; the already observed phenomenology remains largely unchanged as $N$ increases, thus corroborating that thermal resonance is not a size effect that would disappear in the thermodynamic limit, as can be appreciated in Fig.~\ref{fig:9}(a) where we present the average heat flux normalized to the system size as a function of the external drive $\omega$ for various $N$ values. From these results it can be noticed that the magnitude of the resonant frequency $\Om$ decreases as $N$ increases. This redshift effect, first reported in Ref.~\cite{Beraha15} for the NN instance of the herein studied system, has been previously explained, based on an analysis of the non-driven FK lattice, by noticing that the thermal response time, i.e., the time scale for the energy to diffuse across the system, depends on the system size as $\tau\sim N^2$ for large enough $N$ values wherewith the FK lattice obeys Fourier's law and the diffusive regime is attained; thus the associated frequency scales as $\omega\sim N^{-2}$~\cite{Li08}. However, in our case ---as well as in the NN case previously studied~\cite{Beraha15}--- we do not obtain either a ballistic scaling $\sim N^{-1}$ for small system sizes or a diffusive one for larger ones as was the case of the coupled dissimilar FK lattices under the influence of two time-shuttling thermal reservoirs~\cite{Ren10}. Rather, in our case we obtain a discontinuous redshift of $\Om$ for an increasing system size, as is clear from the results presented in Fig.~\ref{fig:9}(b) ---at least for the $N$ values so far studied---. It can be readily appreciated that the resonant frequency locks in two specific values for two clearly defined intervals of $N$ values for the considered system sizes so far studied. This phenomenology could be the result of a complicated competition of time scales, since there are three involved: ballistic, diffusive, and external driving time scales. It deserves special mention the seemingly power law dependence $\Jm\sim N^\alpha$ with $\alpha\approx0.8$ that can be inferred from the data reported in the first panel of the figure, which is quite remarkable considering that, for the undriven homogeneous FK lattice, $J\sim1/N$ holds. The herein reported anomalous transport behavior is akin to that observed in nonlinear oscillator lattices and carbon nanotubes~\cite{Li05a}, where its complete explanation in terms of the involved heat carriers ---phonons, solitons, and/or breathers--- is still a heavily debated subject~\cite{Nianbei10,Likhachev11}. Therefore the fact that the external drive is able to significantly alter the heat conduction properties of the system could be a signal that another heat carrier other than the interacting phonons so far considered is responsible of the aforementioned power law dependence. Although a corroboration of these hypothesis lies outside of the scope of the present study, we do believe that it is worthwhile pursuing it in the near future.

\section{Concluding remarks\label{sec:Disc}}

We have investigated the influence of NNN interactions on the thermal resonance phenomenon present in a 1D system composed of two dissimilar FK lattices connected by a weak harmonic coupling on which an external periodic modulation is exerted. This effect could provide a mechanism to deal with the large amounts of heat transported through a molecular junction, which can compromise its integrity~\cite{Wang07b}. Moreover, the NNN interactions afford a first step to consider the effects on thermal resonance of the long-range interactions, such as of the Morse and Lennard-Jones type, present in molecular junctions consisting of polyethylene polymer chains placed between and attached to metal substrates subject to a static thermal bias~\cite{Dinpajooh22}. It is natural to expect, and our results are to a large extent a confirmation, that NNN interactions increase the coupling between both FK lattices, which begin to act as a whole; thus, the magnitude of the resulting thermal resonance is reduced. And indeed, when the asymmetry of the system is reduced, a decrease of the resonant heat flux is observed. However, the latter can attain significant values for highly asymmetric systems. Furthermore, if the magnitude of the harmonic coupling is increased with a fixed asymmetry value, first a significant increment in the heat flux on both sides of the system is obtained and then, beyond certain value, the magnitude of the flux into the colder reservoir is saturated whereas that of the flux towards the hot one experiences a modest increment as the value of the harmonic coupling further increases. Nevertheless, for this value range of the harmonic coupling the magnitude of both fluxes is way higher ---an order of magnitude in some cases--- than those obtained in the corresponding system with purely NN interactions~\cite{Romero20}. This result exemplifies the complex interaction between the structural properties of the lattice and the external driving, which makes difficult to predict the behavior of the system under a specific condition.

Also, the heat transport properties of the driven lattice seem to be different to those of the undriven one, being neither completely ballistic nor diffusive as deduced from the dependence of the resonant frequency on the system size. This last result can be relevant because, although much information on heat diffusion in oscillator lattices is already known~\cite{Zhao06} ---even for ones with NNN~\cite{Xiong17} as well as long-range interactions~\cite{Bagchi17}---, there is a lack of information about the heat diffusion under the influence of an external driving, which is a condition that is more likely to be important in future technological applications. 

There is also a number of possible open questions that we believe are important for this and similar models but that have not been addressed in this paper. This includes analyzing the temperature dependence of the thermal resonance phenomenon, the generality of these results by studying other classes of interaction and/or onsite potentials, the existence of solitons~\cite{Nianbei10,Ming18}, the effect of discrete breathers that could affect heat transport~\cite{Gendelman00,Giardina00,Xiong12,Xiong14}, extension to higher dimensions, etc. We plan to address these issues in the future. Of course, any analytical result concerning thermal resonance would be highly desirable to obtain. We hope that the herein presented results will encourage further investigations on the heat transport properties of systems affected by an external driving in the future. 

\ack
M.~R.~B. thanks Teodoro Hern\'andez del Angel for insightful comments and discussions and Consejo Nacional de Ciencia y Tecnolog\'\i a (CONACYT) Mexico for financial support. A.~G.~M.~R. thanks ``Programa Institucional de Formaci\'on de Investigadores" I.P.N. M\'exico for financial support.


\section*{References}

\providecommand{\newblock}{}

\end{document}